\documentclass[aps,prl,twocolumn,superscriptaddress]{revtex4}

\usepackage{graphicx}
\usepackage{subfigure}

\begin{document}


\title{Imperfect Imitation Can Enhance Cooperation}


\author{Carlos P.\ Roca}
\email[]{cproca@math.uc3m.es}
\homepage[]{http://www.gisc.es}
\affiliation{GISC/Departamento de Matem\'aticas, Universidad Carlos III de
Madrid, Spain}

\author{Jos\'e A.\ Cuesta}
\affiliation{GISC/Departamento de Matem\'aticas, Universidad Carlos III de
Madrid, Spain}

\author{Angel S\'anchez}
\affiliation{GISC/Departamento de Matem\'aticas, Universidad Carlos III de
Madrid, Spain}
\affiliation{Instituto de Biocomputaci\'on y F\'isica de Sistemas Complejos
(BIFI), Universidad de Zaragoza, Spain}
\affiliation{Instituto de Ciencias Matem\'aticas CSIC--UAM--UC3M--UCM,
Madrid, Spain}



\begin{abstract}
The promotion of cooperation on spatial lattices is an important issue
in evolutionary game theory. This effect clearly depends on the update
rule: it diminishes with stochastic imitative rules whereas it increases with
unconditional imitation. To study the transition between both regimes, we
propose a new evolutionary rule, which stochastically combines unconditional
imitation with another imitative rule. We find that, surprinsingly, in
many social dilemmas this rule yields higher cooperative levels than
any of the two original ones. This nontrivial effect occurs because
the basic rules induce a separation of timescales in the microscopic processes
at cluster interfaces. The result is robust in the space of $2 \times 2$
symmetric games, on regular lattices and on scale-free networks.
\end{abstract}



\maketitle


Why individuals cooperate is a key problem in a wide range of disciplines
\cite{pennisi:2005}, being studied theoretically mainly within the framework of
evolutionary game theory \cite{gintis:2000}. One of the proposed mechanisms to
explain cooperation is network reciprocity \cite{nowak:2006b}, and so different
population structures are known to have an influence on the evolutionary outcome
of social dilemmas \cite{szabo:2007}. All these models incorporate some kind of
evolutionary dynamics \cite{hofbauer:1998}, whose update rules may play a
crucial role in the results. For example, the well-known promotion of
cooperation in Prisoner's Dilemma enforced by spatial lattices \cite{nowak:1992}
is linked to a particular non-stochastic rule (unconditional imitation), and
this effect is greatly reduced if another dynamics, imitative but stochastic, is
employed \cite{roca:2008}.

Imitation is a well-known feature of human behavior
\cite{pingle:1996,apesteguia:2006}. By an imitative dynamics we understand an
update rule that makes individuals copy, within certain constraints, the
strategy of those other players that are doing better, or, in game theoretical
terms, that are obtaining higher payoffs from the game. In the case of network
reciprocity, the range of individuals that every player takes into account is
limited to her nearest neighbors on the network. Two
of the most frequently used imitative dynamics in the literature are  the
\emph{unconditional imitation rule} and the \emph{replicator
rule} \cite{szabo:2007}. In the former, individuals acquire the strategy of the
player with the maximum payoff in their neighborhood including themselves
\cite{nowak:1992}. In the latter, players choose a neighbor at random and copy
her strategy with probability proportional to the difference of payoffs,
provided the neighbor's payoff is greater than hers. It can be proven
\cite{helbing:1992b} that in large well-mixed populations this last rule induces
an evolutionary dynamics equal to the replicator equation, thus leading the
population to asymptotic states very closely related to the evolutionary stable
equilibria of the game \cite{hofbauer:1998}. In structured populations,
though, the evolutionary outcome may greatly differ from the equilibria of the
game \cite{szabo:2007}, and different evolutionary dynamics can yield very
different results, as in the example above. Therefore, it is very
relevant to study the dependence of the promotion of cooperation on the
evolutionary rules and their robustness against perturbations.

In this work, we focus on this issue within the framework of unconditional
imitation and spatial lattices. It may be argued that sometimes individuals can
be able to identify the strategy of all their neighbors and correctly assess
their earnings, but it is difficult to assume that all this complex process may
proceed without errors or disturbances. Indeed, previous work has
pursued this enquiry, using a Moran-like rule with weighted probabilities
\cite{nowak:1994}, finding a progressive lowering of the cooperation levels
as the rule differs from unconditional imitation. Here we use a
different approach, which consists in stochastically combining unconditional
imitation with another less demanding imitative rule. Thus, when an individual
is to update her strategy, she follows unconditional imitation with probability
$1\!-\!\rho$ and the other rule with probability $\rho$ (the other rule is
the replicator rule, unless stated otherwise). The resulting evolutionary
rule, which we call the \emph{$\rho$-rule} in the following, is local and
imitative, and the parameter $\rho \in [0,1]$ measures the perturbation
introduced. Notice that this setting differs from a mutation scheme, as the
players do not acquire indefinitely, or until the next mutation, the secondary
rule.

We have studied computationally the outcome of the $\rho$-rule with $2
\times 2$ symmetric games, which are games with 2 players
who choose between 2 strategies, with no difference in role. We use the
following parametrization of the payoff matrix \cite{santos:2006a,roca:2008}
\begin{equation}
\label{eq:payoff-matrix}
\begin{array}{cc}
  & \begin{array} {cc} \mbox{C} & \mbox{D} \end{array} \\
  \begin{array}{c} \mbox{C} \\ \mbox{D} \end{array} &
  \left( \begin{array}{cc} 1 & S \\ T & 0 \end{array} \right),
\end{array}
\end{equation}
where rows represent the strategy (C for cooperate and D for defect) of the
player who obtains the payoff and columns that of her opponent.
Restricting parameters to the square $-1 < S < 1$, $0 < T < 2$, we
have the Harmony game \cite{licht:1999} ($0 < S,\, T < 1$) and three classic
social dilemmas: the Prisoner's Dilemma \cite{axelrod:1981} ($-1 < S <
0,\, 1 < T < 2$), the Stag Hunt game \cite{skyrms:2003} ($-1 < S
< 0 < T < 1$), and the Hawk-Dove \cite{maynard-smith:1973} or
Snowdrift game \cite{sugden:2004} ($0 < S < 1 < T < 2$). Therefore each game
corresponds to a unit square in the $ST$-plane. We have considered square
lattices with 4- and 8-neighborhoods, doing the update synchronously (all
players play and then they simultaneously update their strategy) or
asynchronously (players play and update their strategy sequentially in random
order).

\begin{figure}
\centering
\includegraphics[width=7cm]{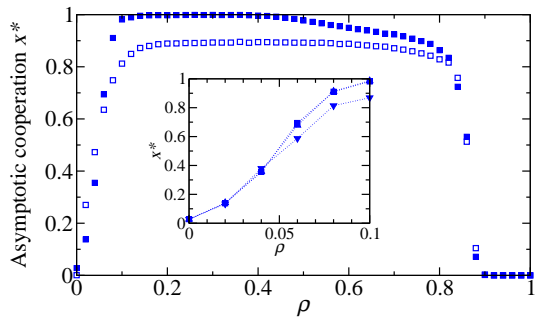}
\caption{Asymptotic density of cooperators $x^*$ in the Prisoner's Dilemma
($S=-0.7$, $T=1.1$), as a function of $\rho$, with
synchronous (filled squares) or asynchronous (empty squares)
update. Population size $N=10^4$ individuals, arranged on a square
lattice with 8 neighbors and periodic boundary conditions. The
initial fraction of cooperators is 0.5, randomly distributed. Simulation time
is $10^4$ generations. The asymptotic values are obtained as the mean over the
last $10^3$ generations, averaged over 100 realizations. Inset: Results with
population sizes $N=2500$ (triangles down), $N=10^4$ (squares) and $N=4.10^4$
(triangles up). The results for $N=10^4$ and $N=4.10^4$ are virtually identical,
and all three are very similar for $\rho > 0.2$. Lines are a guide to the eye.}
\label{fig:pd}
\end{figure}

Figure~\ref{fig:pd} displays an example of the nontrivial behavior obtained with
the $\rho$-rule, showing the asymptotic fraction of cooperators
$x^*$ as a function of the probability $\rho$, for a Prisoner's Dilemma of
parameters $S=-0.7$ and $T=1.1$. With this game, both $\rho=0$ (unconditional
imitation) and $\rho=1$ (replicator rule) result in full defection, but a large
range of values of $\rho$ yield almost full cooperation. This phenomenon
resembles resonance-like behavior found in other game-theoretical models
\cite{szabo:2005a,helbing:2009}, although in this case the effect is
very large and occurs for a wide range of $\rho$.

\begin{figure*}
\centering
\includegraphics[width=17cm]{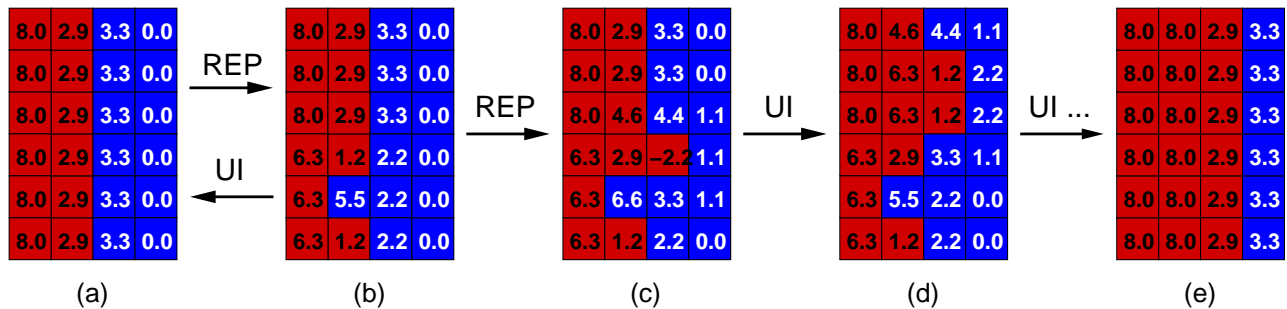}
\caption{Flat interface evolution, with synchronous update, in the Prisoner's
Dilemma of Fig.~\ref{fig:pd} and $\rho>0$. Each position shows the strategy
and payoff of a player. Cooperators are depicted in red and defectors in
blue. Changes of strategy made under unconditional imitation are labeled with
UI, and those under the replicator rule with REP.}
\label{fig:iface}
\end{figure*}

The explanation of this counter-intuitive result lies in the
different microscopic dynamical processes, with greatly differing timescales,
that the update rules induce, something that is known to sometimes stabilize
transients yielding paradoxical outcomes, like in the famous Parrondo's
games \cite{harmer:1999,parrondo:2000,fletcher:2004}. In this case, both rules
enhance cooperation by means of the formation and growth of clusters of
cooperators \cite{nowak:1992}. With unconditional imitation clusters grow
rapidly and mostly with flat interfaces, whereas with the replicator rule
clusters grow much more slowly and have much rougher interfaces
\cite{roca:2008}. Let us consider a flat interface of opposing cooperators and
defectors in a 8-neighbor lattice, as depicted in Fig.~\ref{fig:iface}(a). With
unconditional imitation ($\rho=0$) defectors at the boundary become cooperators
if $5+3S > 3T$. Otherwise the interface remains frozen, and clusters of
cooperators are not able to grow. With the replicator rule ($\rho=1$),
cooperators at the boundary become, with a certain probability, defectors and
the interface roughens progressively. When $\rho>0$ the flat interface slowly
becomes rougher because of those players that happen to follow the replicator
rule, and then nearby defectors that happen to follow unconditional imitation
rapidly become cooperators precisely because of the irregularities at the
interface. For example, starting from the flat interface in
Fig.~\ref{fig:iface}(a) a cooperator will use the replicator rule
and compare payoff with one of the opposing defectors with probability $3 \rho /
8$, becoming with some probability a defector [Fig.~\ref{fig:iface}(b)].
If this player follows unconditional imitation in the next update she
will switch back to cooperation. If not, with a certain probability, one of
the nearby defectors becomes herself a cooperator, using again the replicator
rule, thus producing the kink of Fig.~\ref{fig:iface}(c). This configuration is
critical because if nearby defectors follow unconditional
imitation in the next time steps, they will immediately become cooperators, and
the interface will advance one step to the right [Figs.~\ref{fig:iface}(d,e)].
Simulations show this kind of growth process, where irregularities at interfaces
of clusters appear in configurations similar to that of Fig.~\ref{fig:iface} and
also in cluster corners, giving rise to cascades of conversions from
defectors to cooperators along the interfaces.

As is typical for dynamics driven by unconditional imitation \cite{roca:2008},
the population ends up in full cooperation as long as there are, at initial
time, small clusters of cooperators that resist defector invasion and
grow. Greater population sizes mean larger probabilities of these clusters to
occur, and so there is a dependence on system size in the region close to
$\rho = 0$ (inset of Fig.~\ref{fig:pd}). Note also that the time of
convergence diverges in the limit $\rho \to 0$, because the interface
instabilities explained above take place with a probability proportional to
$\rho^2$.

\begin{figure*}
\centering
\includegraphics[width=18cm]{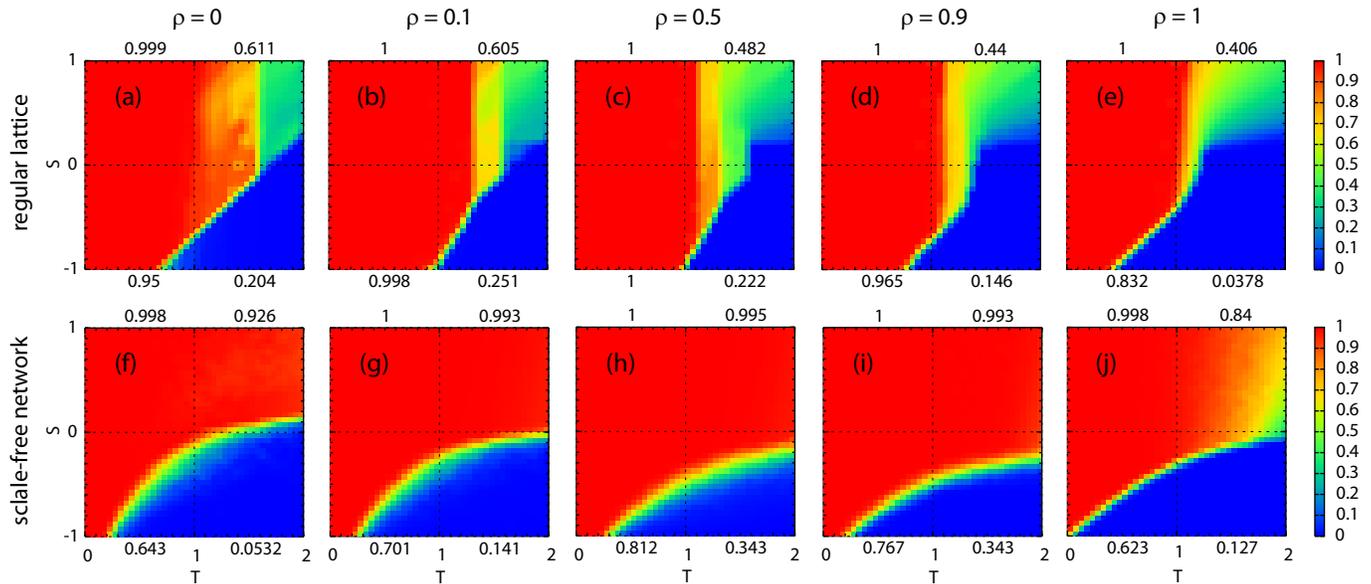}
\caption{Asymptotic fraction of cooperators, for different social dilemmas
defined by $S$ and $T$, as $\rho$ varies from 0 to 1. The population is
arranged on a square lattice [first row, panels (a) to (e)] or a
Barab\'asi-Albert scale-free network [second row, panels (f) to (j)], both of
mean degree equal to 8. The update is synchronous. Remaining parameters are as
in Fig.~\ref{fig:pd}.}
\label{fig:st}
\end{figure*}

The fact that this mechanism is rooted in the basic microscopic processes that
the update rules cause is a hint that it should occur with other games where
these processes are also known to take place \cite{nowak:1992,roca:2008}.
Figures~\ref{fig:st}(a-e) illustrate this point, displaying
the asymptotic fraction of cooperators in the same network topology than
Fig.~\ref{fig:pd}, for different values of $\rho$ and in the games
introduced above: Harmony (upper left square), Stag Hunt (lower left), Snowdrift
(upper right) and Prisoner's Dilemma (lower right).
To have a measure of the global influence on each kind of game, each square
displays the average value of the asymptotic fraction of cooperators achieved in
it. Notice how the transition between full cooperation and full defection of
Fig.~\ref{fig:st}(a) advances in the Prisoner's Dilemma square for
intermediate values of $\rho$. With unconditional imitation ($\rho=0$) this
boundary is given by the payoff equality between cooperators and defectors at
both sides of a flat interface \cite{nowak:1992}, which for the setting of
Fig.~\ref{fig:st}(a) corresponds to  $T-S=5/3$. With $0<\rho \ll 1$
[Figs.~\ref{fig:st}(b,c)], this boundary is instead given by $2T-S=3$, which is
precisely the condition for the payoff equality of the players who determine the
start of cascades at the interfaces, namely the players on third row, second and
third columns, of Fig.~\ref{fig:iface}(c). Note also that the transition line at
$T=8/5$, which determines the instability of inwards corners of cooperators
\cite{nowak:1992}, is preserved for $\rho>0$, and that a new line appears at
$T=4/3$, which gives the condition under which the irregularities of cluster
interfaces, caused by the replicator rule, trigger a defector invasion under
unconditional imitation (compare the payoff of the defector at fifth row and
second column of Fig.~\ref{fig:iface}(c) with that of a cooperator two positions
to the left).

Considering other stochastic imitative rules to be used in the $\rho$-rule, like
the multiple replicator or the Moran rules \cite{szabo:2007,roca:2008}, does not
significantly change the results. In fact, any imitative local rule that
destabilizes the interfaces of clusters of cooperators and that works on a
slower timescale than unconditional imitation (as expected for stochastic
rules) will produce qualitatively similar results. We have found this
phenomenon even with a \emph{random local rule} (players just adopt the
strategy of one randomly chosen neighbor), providing that $\rho \ll 1$. The
effect is also robust against changes in the degree of the network and is found
with synchronous and asynchronous update.

Another important variation to consider in this model is that of the topology of
the underlying network. Apart from the clustering of spatial lattices
\cite{nowak:1992}, the degree heterogeneity of scale-free networks is another
topological property known to have an important impact on the evolution of
cooperation \cite{santos:2006a}. In consequence, we have also studied the
evolutionary outcome that the $\rho$-rule yields on populations structured
according to Barab\'asi-Albert scale-free networks \cite{albert:2002}. The
results are presented in Figs.~\ref{fig:st}(f-j).

\begin{figure}
\centering
\includegraphics[width=7cm]{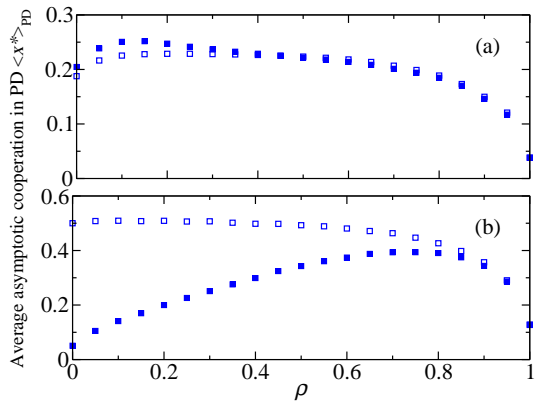}
\caption{Average asymptotic cooperation in the Prisoner's
Dilemma $\langle x^* \rangle_{\mathrm{PD}}$, as a
function of $\rho$.
Populations are distributed on square lattices (a) or scale-free networks (b),
both of mean degree = 8. The update is synchronous (filled squares) or
asynchronous (empty squares). Remaining parameters are as in
Fig.~\ref{fig:pd}. Notice the different scales of
$\langle x^* \rangle_{\mathrm{PD}}$ in (a) and (b).}
\label{fig:pd-rho}
\end{figure}

From the point of view of the evolution of cooperation Fig.~\ref{fig:st}
shows that, for a large range of $\rho$, practically full cooperation is
obtained in Stag Hunt (Snowdrift) games on spatial (scale-free) networks. Very
importantly, with the $\rho$-rule the population achieves full cooperation
precisely in those games where the underlying population structure has
its greatest impact. Regarding the Prisoner's Dilemma, we have studied the
average asymptotic cooperation $\langle x^* \rangle_{\mathrm{PD}}$
in the corresponding unit square, obtaining the results shown in
Fig.~\ref{fig:pd-rho}, both for spatial lattices (a) and scale-free networks
(b). Interestingly, some differences appear depending on the synchronicity of
the update. With synchronous update, a maximum average cooperation is achieved
for an optimum $\rho$, whereas for asynchronous update the result is rather a
plateau over a large range of $\rho$. On spatial lattices the differences are
small, but on scale-free networks they are strikingly large, specially for
$\rho \ll 1$. We have shown elsewhere \cite{roca:2008} that, with spatial
lattices, unconditional imitation is the only rule that yields different results
depending on the synchronicity of the update, but to our knowledge this large
sensitivity in the case of scale-free networks has not been reported
in the literature and so it deserves further investigation.

In conclusion, we have introduced an evolutionary rule that allows to relax the
demanding requirements of the unconditional imitation rule, while maintaining
the basic properties of imitative behavior and local information. We have found
that, for a wide range of $\rho$, the cooperation levels that it yields are not
only preserved but in many cases they are even enhanced. This conclusion is
general and independent of the details of the model; hence the reported
mechanism might have an impact on many other evolutionary games. This work
thus offers a new perspective on the significance of imitative dynamics, in
general, and unconditional imitation, in particular: Instead of considering the
promotion of cooperation adscribed to this rule as a singularity, we can now see
it as a more robust outcome. Finally, and from a more general viewpoint, this
result belongs to a wider class of paradoxical behaviors originated in the
stabilization of transients by the combination of two dynamics, Parrondo's
paradox being the best known example
\cite{harmer:1999,parrondo:2000,fletcher:2004}.

\begin{acknowledgments}
We thank G. Szab\'o for a critical reading of the manuscript.
This work is supported by MICINN (Spain) under Grants
Ingenio-MATHEMATICA and MOSAICO, and by Comunidad de Madrid (Spain) under Grants
SIMUMAT-CM and MOSSNOHO-CM.
\end{acknowledgments}

\bibliography{evol-coop}

\end{document}